\documentclass[aps,twocolumn,showpacs]{revtex4}

\usepackage{graphicx}
 
\newcommand{\be}{\begin{equation}} 
\newcommand{\ee}{\end{equation}} 
\newcommand{\ba}{\begin{eqnarray}} 
\newcommand{\ea}{\end{eqnarray}} 
\newcommand{\sighs}{\sigma_{\rm hs}} 
\newcommand{\sigwca}{\sigma_{\tt WCA}} 
\newcommand{\sigbh}{\sigma_{\tt BH}} 
 
\def\diff{\mathop{\rm\mathstrut d\!}\nolimits} 
\def\C60{C$_{60}$} 
 
\begin{document} 
 
\title{Theoretical description of phase coexistence in model \C60} 
\author{D.~Costa\footnote{Corresponding author, e-mail: 
{\tt costa@tritone.unime.it}}, G.~Pellicane, and C.~Caccamo}
\affiliation{Istituto Nazionale per la Fisica della Materia (INFM) and 
Dipartimento di Fisica \\Universit\`a di Messina, 
Contrada Papardo, C.P. 50, 98166 Messina -- Italy} 
\author{E.~Sch\"oll-Paschinger and G.~Kahl} 
\affiliation{Institut f\"ur Theoretische Physik and CMS, TU Wien \\ 
Wiedner Hauptstra$\beta$e 8-10, A-1040 Wien -- Austria} 
 
\begin{abstract} 
We have investigated the phase diagram of a pair interaction 
model of 
\C60 fullerene~[L.~A.~Girifalco, J. Phys. Chem. {\bf 96}, 858 (1992)], 
in the framework provided by 
two integral equation theories of the liquid state, namely
the Modified Hypernetted Chain 
(MHNC) implemented under
 a global thermodynamic consistency constraint, and the 
Self-Consistent Ornstein-Zernike Approximation~(SCOZA),
and by a Perturbation Theory~(PT) with various degrees of refinement, 
for the free energy of the solid phase.
 
We present an extended assessment of such theories as set against 
a recent Monte Carlo study
of the same model 
[D.~Costa, G.~Pellicane, C.~Caccamo, and M.~C.~Abramo, 
J. Chem. Phys. {\bf 118}, 304 (2003)]. 
We have compared the theoretical 
predictions with the corresponding simulation results 
for several thermodynamic properties like 
the free energy, the pressure, and the internal energy. 
Then we  have
determined the 
phase diagram of the model, 
by using either the  SCOZA, or the MHNC, or the PT 
predictions for one of the coexisting phases, and the 
simulation data for the other phase, in order to separately
ascertain the accuracy of each theory. 
It turns out that the overall appearance of  
the phase portrait is 
reproduced fairly well by all theories, with remarkable accuracy 
as for the melting line and the solid-vapor equilibrium.
All theories
show a more or less pronounced discrepancy with
the simulated
fluid-solid coexistence pressure, above the triple point.
The MHNC and SCOZA results  
for the liquid-vapor coexistence, as well as for the 
corresponding critical points, are quite accurate;
the SCOZA tends to underestimate the density corresponding to 
the freezing line. All results are discussed in terms 
of the basic assumptions underlying each theory. 

We have selected the MHNC for the 
fluid and the first-order PT 
for the solid phase,  as the most accurate  tools to 
investigate the phase behavior of the model
in terms of 
 purely theoretical approaches.
It emerges that 
the use of different procedures 
to characterize 
the fluid and the solid phases
provides  a semiquantitative
reproduction of  the thermodynamic
properties 
of the \C60 model at issue.
 
The overall results appear as
a robust benchmark  
for further theoretical investigations on higher order  
C$_{n>60}$ fullerenes, as well as on other fullerene-related materials, 
whose description can be
based on a modelization similar to that adopted in this work. 
 

\end{abstract} 

\maketitle 

\columnseprule 0pt
 
\section{Introduction} 
 
A current model for \C60  
is based on a ``smeared out'' spherical representation of the 
fullerene molecules, early proposed by Girifalco~\cite{Girifalco},  
where the carbon atoms give rise to a uniform interaction distributed  
over the molecular cage surface. 
A straightforward integration leads to an analytical 
central pair potential, characterized by a harsh repulsive core 
followed by a deep attractive well, 
which rapidly decays with the interparticle distance. 
 
A detailed 
characterization of the thermodynamic properties of
the Girifalco model
is of relevant interest in several respects: for instance, 
first-principle studies of 
the \C60 interaction give results 
very similar to those predicted 
through the Girifalco model~\cite{Pacheco,Silva1};  
the latter has been widely used to model  
C$_{n>60}$ fullerenes with 
$n=70$, 76, and 84 (see~\cite{Abramo,Micali,Silva,Kahl} and references  
cited therein), as well as 
other hollow nanoparticles like carbon onions 
and  metal dichalcogenides like GaAs and CdSe
(also termed inorganic fullerenes)~\cite{Safran}. 
Umiguchi and coworkers~\cite{Hasegawa0} proposed 
a semiempirical modification of the Girifalco interaction  
to account for the behavior of  solid \C60 at  
low temperature; such  
a generalization might prove useful for the 
analysis of the lattice properties under impurity doping. 
 
As far as a more speculative analysis of the Girifalco model is concerned, 
several intriguing aspects  
are related to the overall appearance of its phase portrait,  
characterized by a narrow liquid pocket that extends  only 
a few tens degrees over 
the triple point temperature~\cite{Hasegawa,noi}. 
This borderline behavior, as for 
the existence of a stable liquid phase,  
basically depends on the interplay 
between entropic demands, imposed by excluded-volume effects 
at short distance,
and energetic contributions, due to the  
attractive part of the interaction potential.
We may recall, 
as an indication of the implied subtleties,  
the early controversy on the location of  
the liquid-vapor critical point~\cite{Mooij,Cheng},  
clarified in successive studies~\cite{Hasegawa2,Fucile},  
and the discrepancies  
on the position of the freezing line 
and of the triple point, related to 
the  procedure adopted  
for the calculation of 
the liquid-solid equilibrium~\cite{Hasegawa,Cheng,CMHNC}. 
Recently (and unexpectedly, on the basis of a previous study 
on the Lennard-Jones fluid~\cite{Silva2}),  
Fartaria and coworkers~\cite{Silva1} have found 
that even the same simulation approach 
leads to distinct predictions on the 
freezing line of the Girifalco model, provided that
different starting thermodynamic conditions are employed.  
 
We have recently 
reconciled such disparate results  
in the context of extensive 
Monte Carlo calculations of the free energies of  
both the fluid and the solid phases of the model~\cite{noi,noi1}.  
We have concluded 
that the solid-fluid equilibrium is strongly affected by 
the deep, short-range attractive well in the interaction potential, 
at variance with systems whose freezing behavior  
is essentially  dominated by steric effects. 
As a consequence,
the freezing transition of the fluid is driven 
to lower densities (and the liquid pocket is reduced accordingly),  
mainly by energetic effects, in agreement with 
a common scenario recently proposed 
for fluids interacting
through short-range forces~\cite{Louis}.
 
The unusual aspects of the phase behavior  
have occasioned several
studies on the \C60 model  
by means of 
refined theoretical tools, like for instance 
integral equation theories for  
the fluid phase~\cite{Cheng,CMHNC},  various
density functional approximations~\cite{Hasegawa2,Mederos,Hasegawa3},  
and the hierarchical reference theory~\cite{HRT}.  
Preliminary studies have been recently carried out, 
based on the Modified Hypernetted  
Chain approach (MHNC,~\cite{MHNC}), solved under 
a global thermodynamic consistency 
constraint~\cite{Peppe},  
and on the Self-Consistent Ornstein-Zernike Approximation 
(SCOZA~\cite{SCOZA,Pin98,Sch02})~\cite{Kahl1}.  
However, such investigations
were often compared to simulation results later revised; 
for example, several authors~\cite{Cheng,CMHNC,Kahl1} have estimated the  
freezing conditions  
on the basis of some well-known one-phase structural 
indicators,  
as the Hansen-Verlet~\cite{Verlet}, or the  
Giaquinta and Giunta~\cite{pvg} 
criteria, whose limited applicability  
in the context of ``energetic'' fluids has been 
discussed in~\cite{noi}. 
 
Hinging on the simulation study of Ref.~\cite{noi}, 
it is  now worth 
reconsidering a detailed analysis of theoretical predictions 
for the phase diagram of the envisaged model.  
Following our preliminary investigations~\cite{Peppe,Kahl1} 
we shall adopt the MHNC and SCOZA theories  
to characterize the thermodynamic properties of the fluid phase. 
As far as the free energy of the solid phase is concerned, 
we use a Perturbation Theory (PT), discussing 
several degrees of refinement, as  fully described in the text. 
Recent studies (see~\cite{Amokrane,Nava} and references)
have demonstrated that
the PT accurately describes the solid phase
and the solid-fluid transition of models characterized
by short-range interactions, as the depletion potentials
resulting from the effective one-component representation
of hard-sphere mixtures.  A second-order expansion  for the 
PT, also analyzed in this work, 
has been used in Ref.~\cite{Foffi}
to investigate the  phase behavior of several 
simple models for globular protein solutions.

This paper is organized as follows: 
the theoretical approaches are described 
in section~II.  Results are reported and  
discussed in Section~III. A short overview of our results
and the conclusions are drawn in Section~IV. 
 
\section{Model and theoretical approaches} 
 
\subsection{The Girifalco model potential} 
\label{sec:2a} 
 
The Girifalco potential is well known in the fullerene literature,  
so we recall only its analytical expression~\cite{Girifalco},  
\ba\label{eq:pot} 
v(r) & = &   -\alpha_1 \left[ \frac{1}{s(s-1)^3} +
                   \frac{1}{s(s+1)^3} - \frac{2}{s^4} \right] \nonumber\\[4pt]
     & & \quad+\alpha_2    \left[ \frac{1}{s(s-1)^9} +
                   \frac{1}{s(s+1)^9} - \frac{2}{s^{10}} \right] \,,
\ea 
where $s=r/d$, $\alpha_1=N^2A/12d^6$, and 
$\alpha_2=N^2B/90d^{12}$; 
$N$ and $d$ are the number of carbon atoms and the diameter, 
respectively, of the fullerene particles, 
$A=32\times10^{-60}$~erg\,cm$^6$ and 
$B=55.77\times10^{-105}$~erg\,cm$^{12}$ are constants entering the 
Lennard-Jones 12-6 potential 
through which two carbon sites on 
different spherical molecules are assumed to interact~\cite{Girifalco}. 
For \C60, $d=0.71$~nm while the node of the
potential~(\ref{eq:pot}),  
the minimum, and its position,
are $r_0\simeq 0.959$~nm,  
$\varepsilon\simeq 0.444\times10^{-12}$~erg 
and $r_{\rm min}=1.005$~nm, respectively. 
 
\subsection{Perturbation theory of the solid phase} 
 
The perturbation approach is based  
on a separation of the 
potential~(\ref{eq:pot}) into  
a reference (purely repulsive) part, $v_{\rm ref}(r)$, and a  
residual, attractive part, $v_{\rm pert}(r)$. 
The reference system is then approximated by 
a solid of hard spheres with the same crystallographic structure, 
by means of a suitable definition 
of the hard-core diameter. 
Given these positions, 
the free energy of the system can be expanded 
around the free energy of the hard-sphere crystal, $F_{\rm hs}$, 
to obtain at first order~\cite{HM}: 
\be\label{eq:exp} 
\frac{\beta F}{N}=\frac{\beta F_{\rm hs}}{N}+ 
\frac{\beta \rho}{2} \int v_{\rm pert}(r) \  
\overline g_{\rm hs}(r) \diff {\mathbf r} \,, 
\ee 
where $N$ is the number of particles, $\beta$
the inverse of the temperature $T$ in units of the Boltzmann constant,
and $\rho$ the number density;
the second term is the thermal average 
of the perturbation energy over the reference system, where 
 $\overline g_{\rm hs}(r)$ is  
the angular average of the pair distribution function
of the hard-sphere solid, 
$\rho^{(2)}({\bf r}_1, {\bf r}_2)$~\cite{Weis,Rascon}. 
 
We have followed the WCA prescription (after  
Weeks, Chandler, and Andersen~\cite{WCA}) to split
the potential~(\ref{eq:pot}), namely: 
\ba\label{eq:wca}
\begin{array}{cc}
v_{\rm ref}(r) = &  
\left\{
 \begin{array}{cc}
 v(r)+\varepsilon    & \quad {\rm if} \ r \le r_{\rm min} \cr
 0                      & \quad {\rm if} \ r > r_{\rm min}
 \end{array}
 \right. \\ \\
v_{\rm pert}(r)= &
\left\{
 \begin{array}{cc}
\quad  -\varepsilon  \quad  &\  \quad {\rm if} \ r \le r_{\rm min} \cr
 v(r)                  & \  \quad {\rm if} \ r > r_{\rm min}
 \end{array}
 \right. \,.
\end{array} 
\ea
The properties of the reference system have been  connected  
to those of the hard-sphere solid through the ``blip function'' 
formalism~\cite{WCA2}, i.e.
the effective hard-sphere 
diameter $\sigwca$ is determined by the implicit relation: 
\be\label{eq:sigma} 
\int y_{\rm hs}(r)  
\{\exp[-\beta v_{\rm pert}(r)]-\exp[-\beta v_{\rm hs}(r)]\}  
\diff {\bf r} = 0\,, 
\ee 
where  $v_{\rm hs}(r)$ is the hard-sphere 
potential (which depends on $\sigwca$) and 
$y_{\rm hs}(r)=g_{\rm hs}(r)\exp[\beta v_{\rm hs}(r)]$  
is the corresponding cavity function.
Another definition of the hard-core diameter (BH,~\cite{BH}),  
namely: 
\be\label{eq:sigma1} 
\sigbh=\int_0^\infty \{1 -\exp[-\beta v_{\rm pert}(r)]\}\diff r \, 
\ee 
has been tested in this work. The expression~(\ref{eq:sigma1}), 
which 
corresponds to a first-order approximation to Eq.~(\ref{eq:sigma})~\cite{HM}, 
is appropriate to describe the rapidly rising 
repulsive interaction $v_{\rm pert}(r)$. 
Other prescriptions for the separation of the  
potential~(\ref{eq:pot}) (see Refs.~\cite{BH} and~\cite{Choi}) 
have been analyzed in this study;  
their predictions for the thermodynamic properties of the model at issue
are, however, 
less accurate on the whole than those obtained 
through the WCA approach and will not be discussed further. 
 
In the results' section 
we also investigate 
a second-order 
correction to the free energy, early 
proposed by Barker and Henderson 
in Ref.~\cite{BH2}, and recently applied to systems with 
short-range interactions by Foffi and coworkers~\cite{Foffi}, namely: 
\ba\label{eq:exp2} 
\frac{\beta F}{N} & =& \frac{\beta F_{\rm hs}}{N}+ 
\frac{\beta \rho}{2} \int v_{\rm pert}(r) \ 
\overline g_{\rm hs}(r) \diff {\mathbf r} \nonumber \\[4pt]  
 & & \quad 
-\frac{\beta \rho}{4} \left(\frac{\partial \rho}{\partial P_{\rm hs}}\right) 
\int v_{\rm pert}(r)^2 \ 
\overline g_{\rm hs}(r) \diff {\mathbf r}\,, 
\ea 
where $P_{\rm hs}$ is the pressure of the solid of hard spheres. 
The correction in Eq.~(\ref{eq:exp2}) was obtained  
by dividing the space into concentric spherical shells, 
and assuming that the volume of each shell 
has the compressibility properties of a macroscopic  
portion of the space (see Ref.~\cite{BH2} for full details 
on the procedure and the approximations involved). 
 
We have resorted to several well-established 
simulation results in order to obtain the properties 
of the  hard-sphere crystal. Specifically, 
 we have used  
the Hall analytical equation of state for the pressure~\cite{Hall}, 
which accurately fits the molecular dynamics data of  
Alder and coworkers~\cite{Alder}: 
\be\label{eq:eoshs} 
\frac{\beta P_{\rm hs}}{\rho} = \frac{3\eta}{\eta_{\rm cp}-\eta} 
+\sum_{\eta=0}^6 a_n \ \gamma^n \, 
\ee 
where $\eta$ is the packing fraction and 
$\eta_{\rm cp}=\pi\sqrt{2}/6$  is
the close packing fraction;  
$\gamma=4(1-\eta/\eta_{\rm cp})$ and  
$a_0=2.557\,696$, $a_1=0.125\,307\,7$, $a_2=0.176\,239\,3$, $a_3=-1.053\,308$, 
$a_4=2.818\,621$, $a_5=-2.921\,934$, $a_6=1.118\,413$. 
The free energy is obtained by thermodynamic integration 
\be\label{eq:fvsrho} 
\frac{\beta F_{\rm hs}(\rho)}{N} = 
\frac{\beta F_{\rm hs}(\overline\rho)}{N} + 
\int_{\overline\rho}^\rho \frac{\beta P_{\rm hs}(\rho')}{\rho'} 
\frac{\diff \rho'}{\rho'} \,, 
\ee 
where for the reference  
free energy, 
$F_{\rm hs}(\overline\rho)$,  we have used 
the Frenkel and Ladd~\cite{Ladd}  Monte Carlo result for the FCC crystal
of hard spheres, namely:
$\beta F^{\rm ex}_{\rm hs}/N =6.537\,9(09)$ 
at $\eta/\eta_{\rm cp}=0.7778$  
in the thermodymamic limit. 
Their determination
followed the Einstein crystal method,
based on the construction of a reversible  
path from the solid under consideration to an Einstein crystal with  
the same crystallographic structure. 
As for the 
radial distribution function, 
$\overline g_{\rm hs}(r)$, we have used 
the analytical form, proposed by Weis~\cite{Weis} and Kincaid  
and Weis~\cite{Kincaid}, which
expresses the distribution as a sum of gaussian 
peaks centered around the nearest-neighbour FCC lattice sites, 
with parameters calculated by  
Choi and coworkers~\cite{Choi} to
fit their simulation results. 
In Ref.~\cite{Choi} an analytical form 
for the cavity function 
$y_{\rm hs}(r)$ inside the hard core  
entering Eq.~(\ref{eq:sigma}) 
is also proposed, by directly 
extending to the solid the Henderson 
and Grundke scheme for the fluid phase~\cite{HG}. 
 
\subsection{Liquid state theories} 
 
We have determined
the thermodynamic and structural properties 
of the fluid region 
through the 
MHNC~\cite{MHNC} and SCOZA~\cite{SCOZA} theories.  
We recall that in the MHNC   
the Ornstein-Zernike equation, 
\be\label{eq:oz} 
h(r) = c(r) + \rho 
\int c(|{\bf r}-{\bf r'}|) h(r') d {\bf r'} \,, 
\ee 
is coupled with the cluster expansion for the 
radial distribution function $g(r)$, namely:
\be\label{eq:mhnc} 
g(r)=\exp [-\beta v(r) + h(r) - c(r) + B(r)] \,, 
\ee 
where $h(r)=[g(r)-1]$ and $c(r)$ are the pair 
and direct correlation functions respectively, 
and $B(r)$ is the bridge function~\cite{HM}. 
A closure to Eqs.~(\ref{eq:oz}) 
and~(\ref{eq:mhnc}) is obtained through the 
Percus-Yevick hard-sphere bridge 
function $B^{\rm PY}(r,\sighs^*)$, with the hard-core diameter 
$\sighs^*$ fixed so
to enforce the thermodynamic consistency of the theory 
(see e.g.~\cite{report}). 
In particular
we require the equality between the fluctuation pressure 
$P_{\rm c}$ and the virial pressure $P_{\rm v}$, namely,  
\be\label{eq:gc1} 
P_{\rm c} \equiv 
\beta^{-1} \int_0^{\rho} \frac{\chi_0}{\chi_{\rm T}} \diff \rho'  
= P_{\rm v} \,. 
\ee 
In Eq.~(\ref{eq:gc1}) $\chi_0$  and $\chi_{\rm T}$ are the
ideal  
gas and the isothermal compressibilities respectively;
the latter is obtained as the 
$q=0$ limit of the Fourier transform of the direct correlation function, 
$\chi_0/\chi_{\rm T}=[1-\rho\tilde c(q=0)]$, 
according to the
fluctuation theory;  the density integral is  
performed at a constant temperature. 
 
The consistency between $P_{\rm v}$
and the pressure estimated via the energy route, 
$P_{\rm e}$,
can be alternatively enforced along isochoric paths, namely: 
\be\label{eq:gc2} 
 P_{\rm e}\equiv \rho^2 \left.  
\frac{\partial F}{\partial \rho} \ \right|_{\rm T} =  P_{\rm v} \,. 
\ee 
In Eq.~(\ref{eq:gc2}) the free energy is obtained 
by integrating the internal energy $U$ 
with respect to the (inverse) temperature at constant density, according 
to the relation: 
\be\label{eq:fvst} 
\frac{\beta F(\rho,T)}{N} = \frac{\beta F(\rho,\overline T)}{N} 
- \int_{\overline T}^T \frac{U(T')}{Nk_BT'} 
\frac{\diff T'}{T'} \,, 
\ee 
where $(\rho,\overline T)$ is a thermodynamic state  
whose absolute free energy $\beta F(\rho,\overline T)/N$ 
is known.  In order 
to avoid to cross the liquid-vapor coexistence region,
combinations of isothermal and isochoric paths 
are required to reach thermodynamic points  
located in the high-density, subcritical 
region of the phase diagram. 
 
As shown in Ref.~\cite{Peppe}, 
the global 
consistency is in general more accurate than the local one;  
the latter procedure
enforces the equality between the
isothermal compressibility calculated via the 
fluctuation theory  on the one hand, and via
the virial route (i.e. by differentiating 
$P_{\rm v}$ with respect to the density) on the other hand. 
The improvement of the global over the local consistency 
approach becomes 
crucial when the range of the 
particle interaction is very short, and/or at high densities 
and low temperatures;
the free energy and the chemical 
potential in particular turn out to be 
almost quantitatively  
predicted~\cite{Peppe}. 
 
As far as the  SCOZA theory 
is concerned, this represents  
an advanced liquid state theory that is based on a mean-spherical~(MSA) 
type closure relation, replacing the prefactor $\beta$ in the closure for 
the direct correlation function by a yet undetermined, state dependent 
function $K(\rho, T)$; this function is fixed by the consistency requirement  
between the energy and the compressibility route 
to thermodynamics~\cite{Pin98,Sch02}. Thus 
for a hard-core potential of diameter $\sighs$  
and with an attractive tail 
$w(r)$, the SCOZA closure relations to the OZ equation  
(\ref{eq:oz}) read: 
\begin{equation}                                 \label{SCOZA_g} 
g(r) = 0 ~~~~~~ r < \sighs 
\end{equation} 
\begin{equation}                                 \label{SCOZA_c} 
c(r) = c_{\rm hs}(r) + K(\rho, T) w(r) ~~~~~~ r \ge \sighs \,. 
\end{equation} 
In relation (\ref{SCOZA_c}) $c_{\rm hs}(r)$ 
 represents the direct correlation  
function for the hard-core system;  
we have used the Waisman parameterization~\cite{Wai73},  
i.e., $c_{\rm hs}(r) = K_0/r \exp[-z_0(r - \sighs)]$  
(with well-described density-dependent coefficients $K_0$ and $z_0$) 
which is known to give accurate results for the hard sphere system. The 
consistency requirement between the two thermodynamic routes mentioned above 
leads to the partial differential equation~(PDE): 
\begin{equation} 
\frac{\partial}{\partial \beta} \left( \frac{1}{\chi_{\rm red}} \right)  
= \rho \frac{\partial^2 u}{\partial \rho^2} \,, 
\end{equation} 
where $\chi_{\rm red}=\chi_{\rm T}/\chi_0$ 
is the reduced (with respect to the ideal gas)  
dimensionless isothermal compressibility given by the
compressiblity route and $u$ is the excess (over ideal 
gas) internal energy per volume provided by the energy route. 
Solution of the above PDE leads to  
$K(\rho, T)$ and thus to the structure and thermodynamics of the system. 
 
The SCOZA has proven to give reliable results for the location of the 
coexistence curve and stays~---~in contrast to most liquid 
state theories~---~reliable even in 
the critical region (see~\cite{Sch02} for an overview).  
Many examples for continuum systems 
as well as for spin systems (in few cases including even binary 
mixtures~\cite{Sch03}) 
have demonstrated its validity. 
Limiting our considerations on
continuum systems, the 
SCOZA suffers from 
two drawbacks (at least in
its present version), despite its many merits: 
as the formalism of the SCOZA largely benefits from the 
availability of the semianalytic solution of the MSA for a hard-core  
Yukawa system (including an arbitrary number of Yukawa tails)~\cite{Arr91},  
its application is limited to systems where 
the attractive part of the interatomic 
potential, $w(r)$,  
can be approximated by a suitable number of Yukawa tails, while
the repulsive core {\it has} to be replaced by a hard-core potential. 
In the present work this has been
done at the node of the potential, $r_0$, as the SCOZA PDE becomes  
unstable for repulsive interactions. While in general the first problem does 
not represent a serious restriction, up to now no remedy has been found to 
include softness of the repulsive part of the interaction.

\section{Results and discussion} 
 
\begin{figure*} 
\begin{center} 
\includegraphics[width=8.5cm,angle=-90]{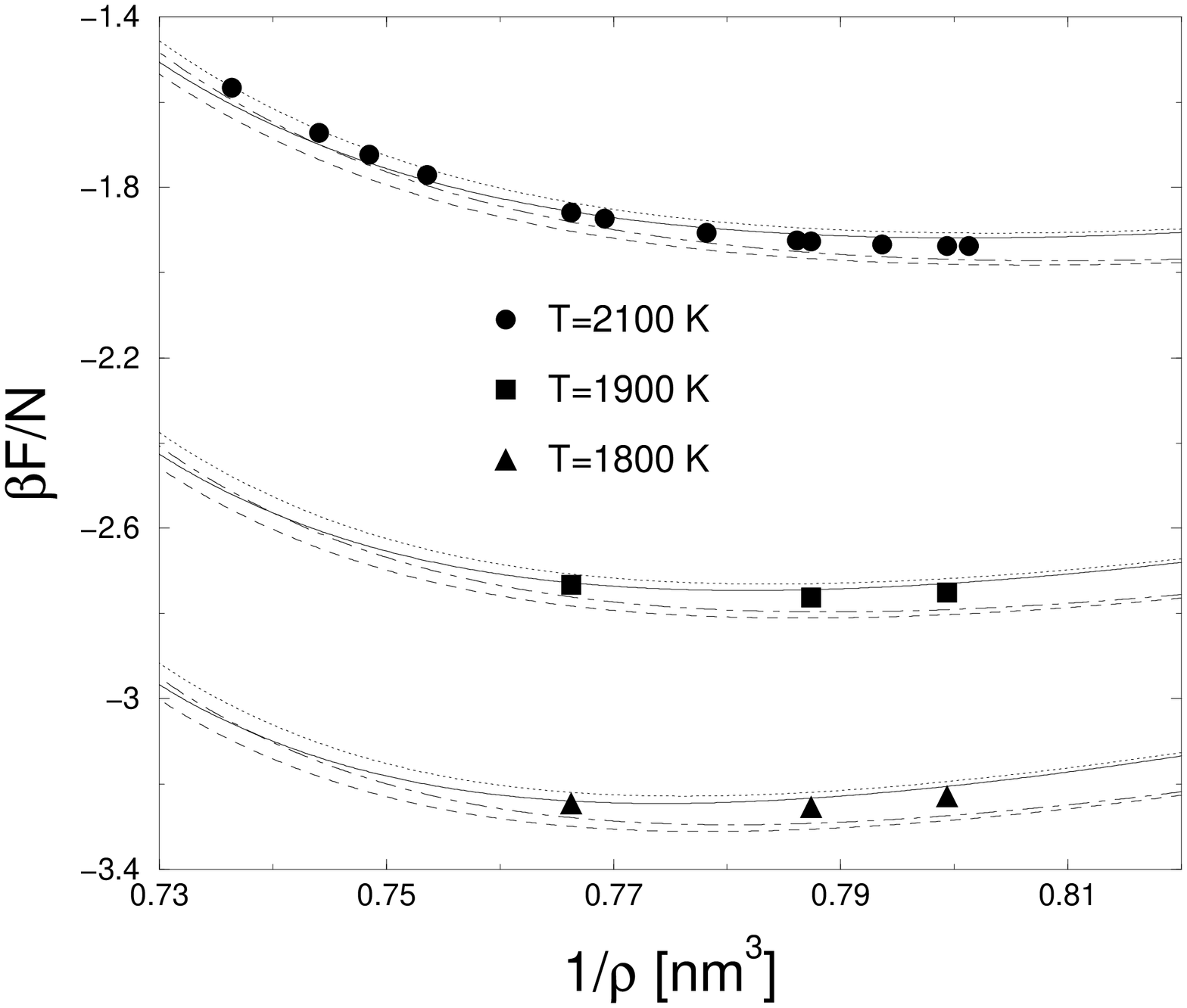} 
\includegraphics[width=8.5cm,angle=-90]{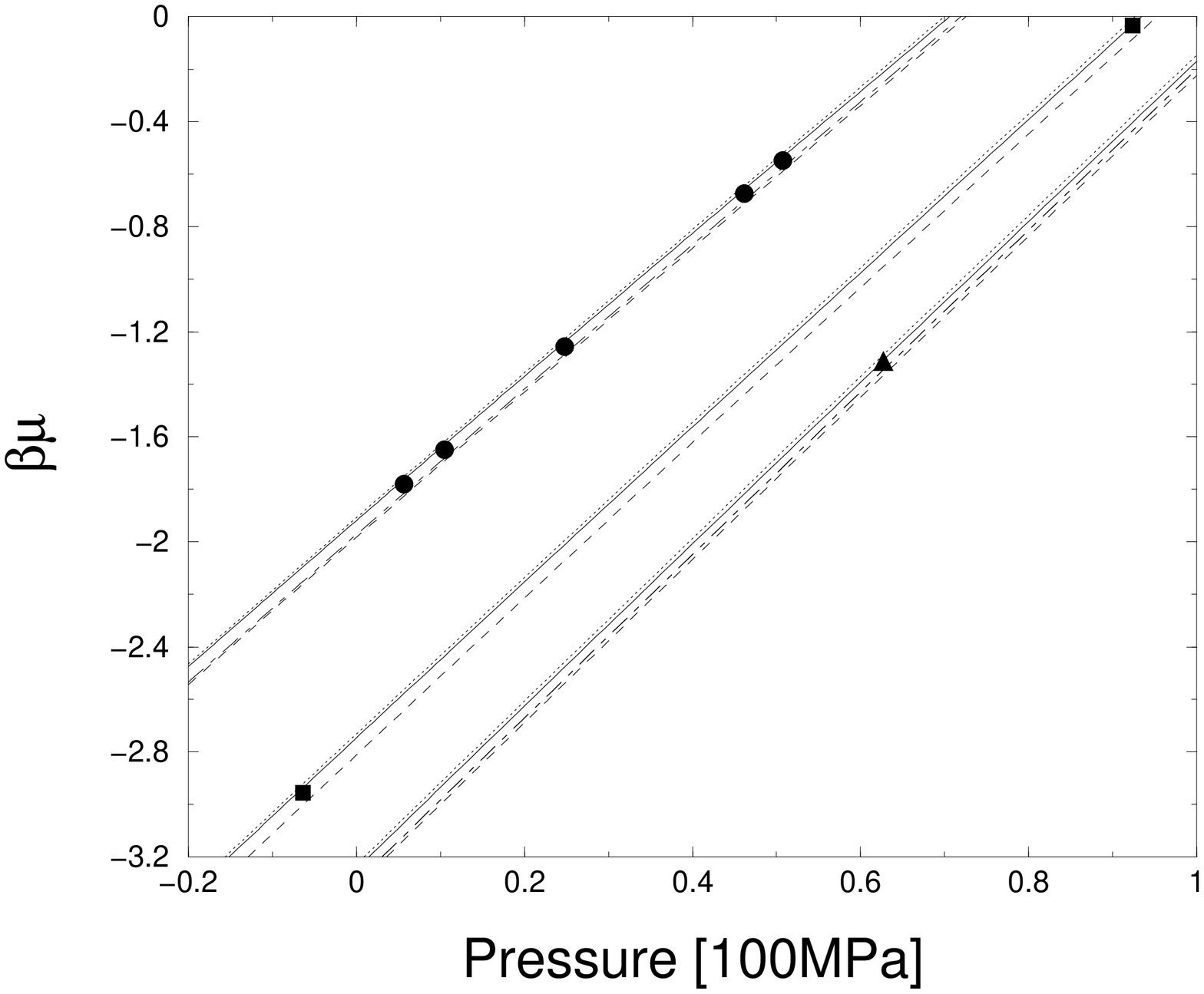}
\caption{PT free energy (top) and chemical potential (bottom)
along three isotherms
in the solid phase:
first-order PT (Eq.~(\ref{eq:exp}))
with $\sigwca$ (Eq.~(\ref{eq:sigma}), dotted line) or $\sigbh$
(Eq.~(\ref{eq:sigma1}), full line),
as equivalent hard-core diameters;
second order PT (Eq.~(\ref{eq:exp2}))
with $\sigwca$ (dot-dashed line) or $\sigbh$ (dashed line).
Symbols represent the  corresponding 
Monte Carlo results~\protect\cite{noi}. 
}\label{fig:freePT} 
\end{center} 
\end{figure*}

\begin{figure*} 
\begin{center} 
\includegraphics[width=9cm,angle=-90]{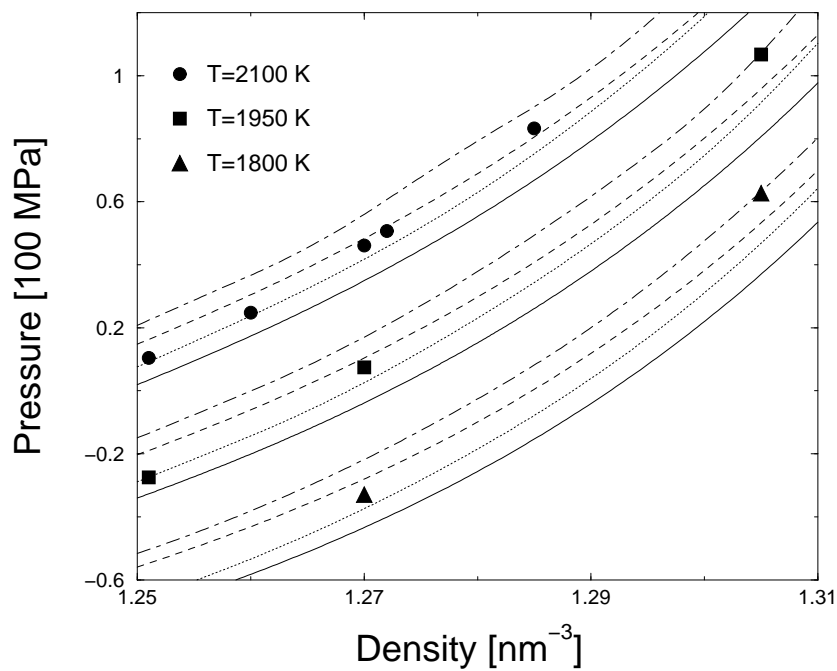} 
\caption{PT equation of state in the solid phase (lines), displayed 
in order of decreasing temperature from top to bottom,
according to the legend of Fig.~\ref{fig:freePT}. 
Symbols: corresponding Monte Carlo results~\protect\cite{noi}.
For clarity sake the isotherms
$T=2100$, 1950, and 1800~K are displayed.
}\label{fig:pressPT} 
\end{center} 
\end{figure*}

\begin{figure*} 
\begin{center} 
\includegraphics[width=9cm,angle=-90]{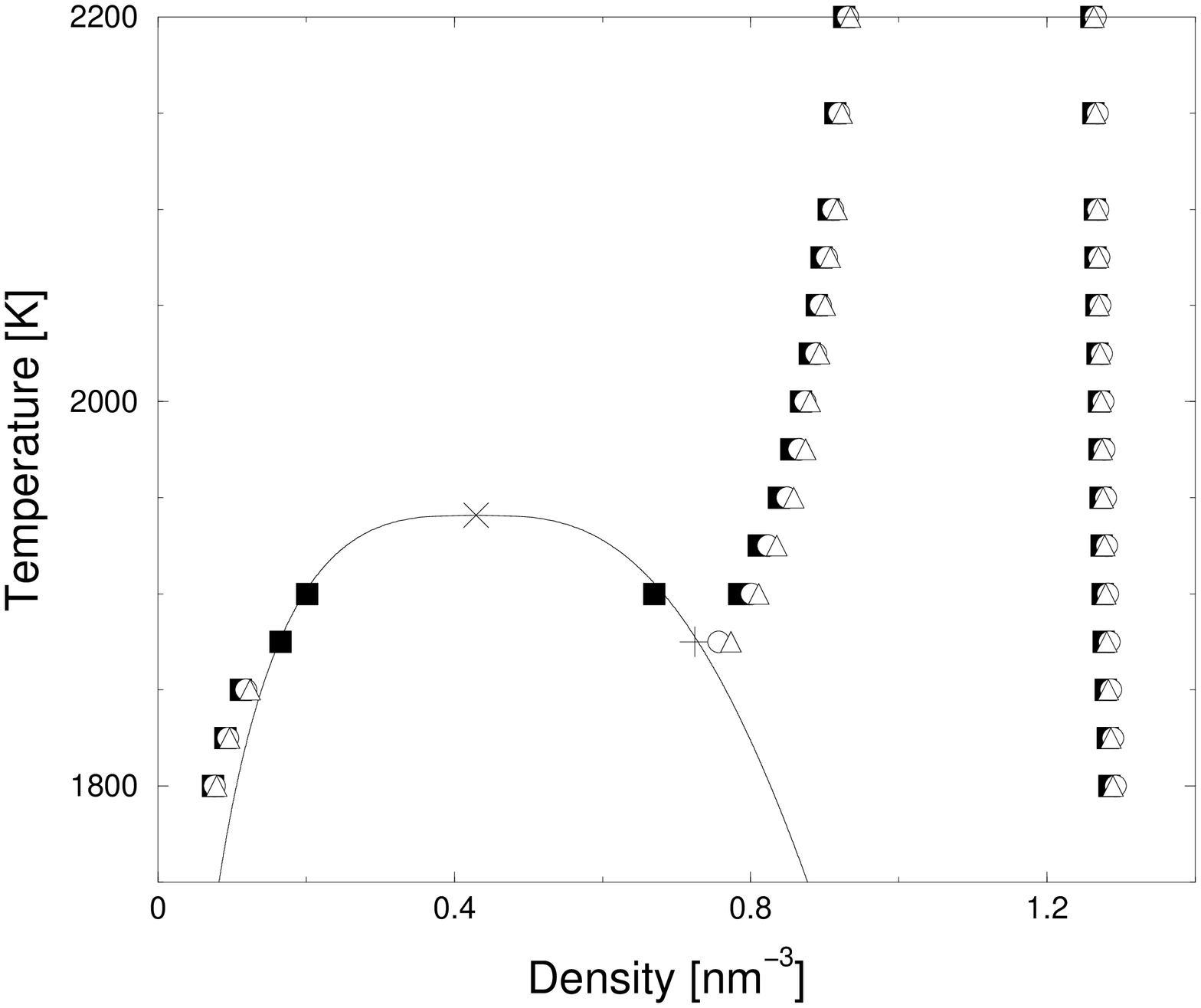}  
\includegraphics[width=9cm,angle=-90]{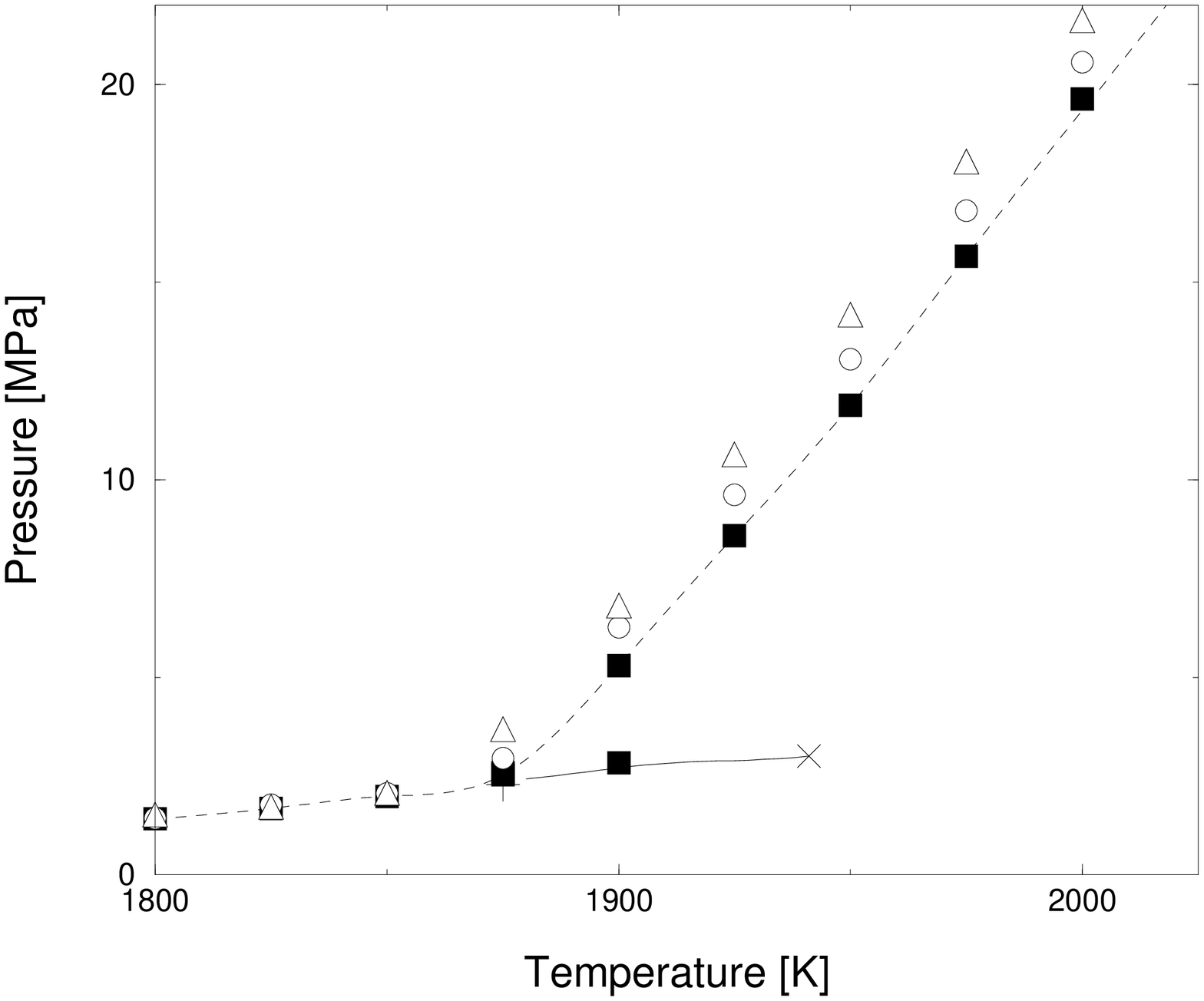} 
\caption{First-order PT fluid-solid coexistence densities (top)
and pressures (bottom) of the \C60 model,
with $\sigwca$ (triangles) or $\sigbh$ (circles) as 
hard-core diameters. 
Squares: Monte Carlo results~\protect\cite{noi}. 
For completeness, 
the GEMC liquid-vapor coexistence densities (top)
and pressures (bottom) are also shown
as full lines~\protect\cite{Fucile}.
The cross and the plus indicate the simulation
critical and triple points, respectively. 
In the bottom panel dashed lines are guides to the eye
for the Monte Carlo results.
}\label{fig:phdiaPT} 
\end{center} 
\end{figure*}

\begin{figure*}[t]
\dimen0=\textwidth 
\advance\dimen0 by -\columnsep 
\divide\dimen0 by 2
\noindent\begin{minipage}[!t]{\dimen0}
\begin{center}
{\includegraphics[width=6.5cm,angle=-90]{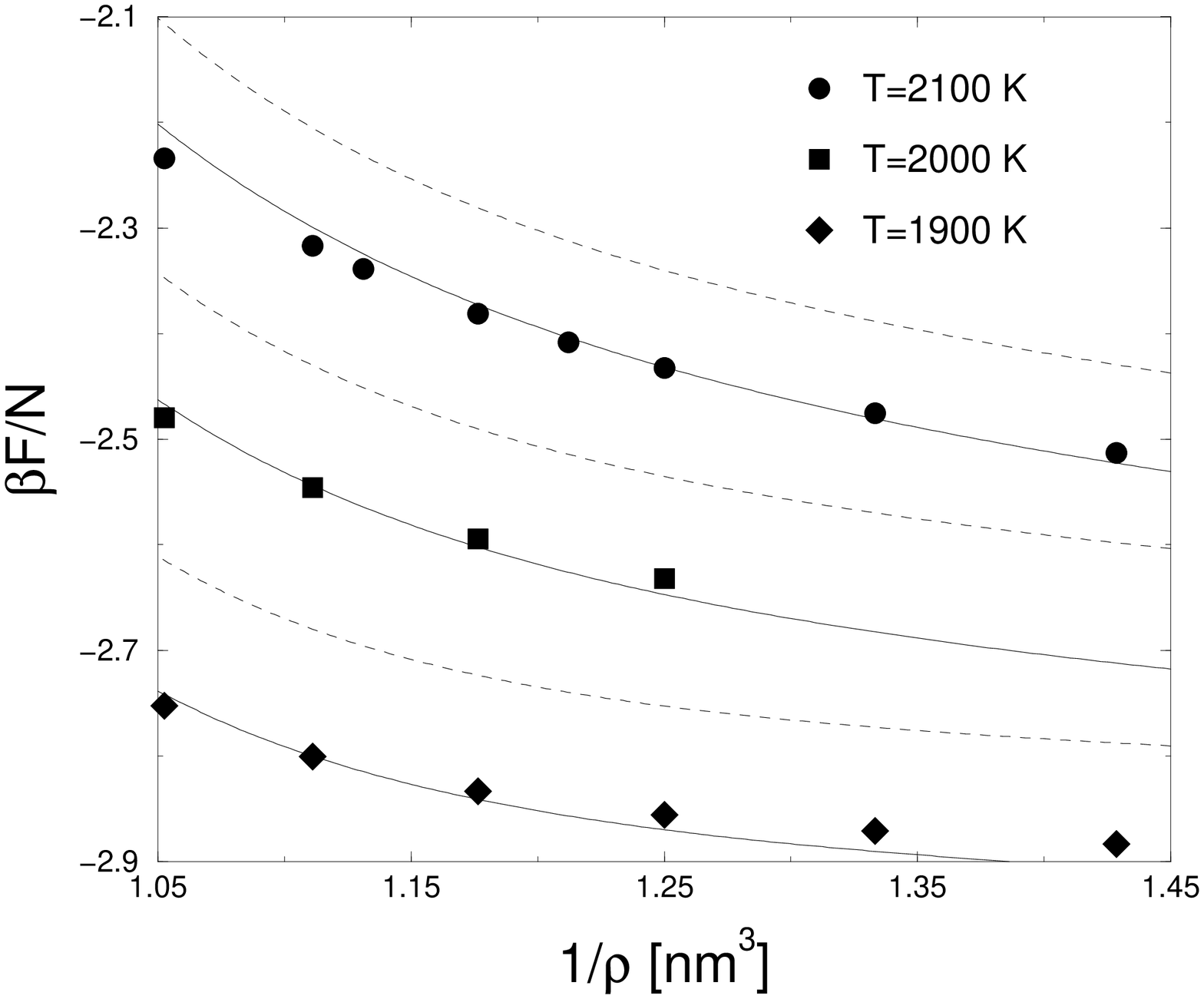}}
{\includegraphics[width=6.5cm,angle=-90]{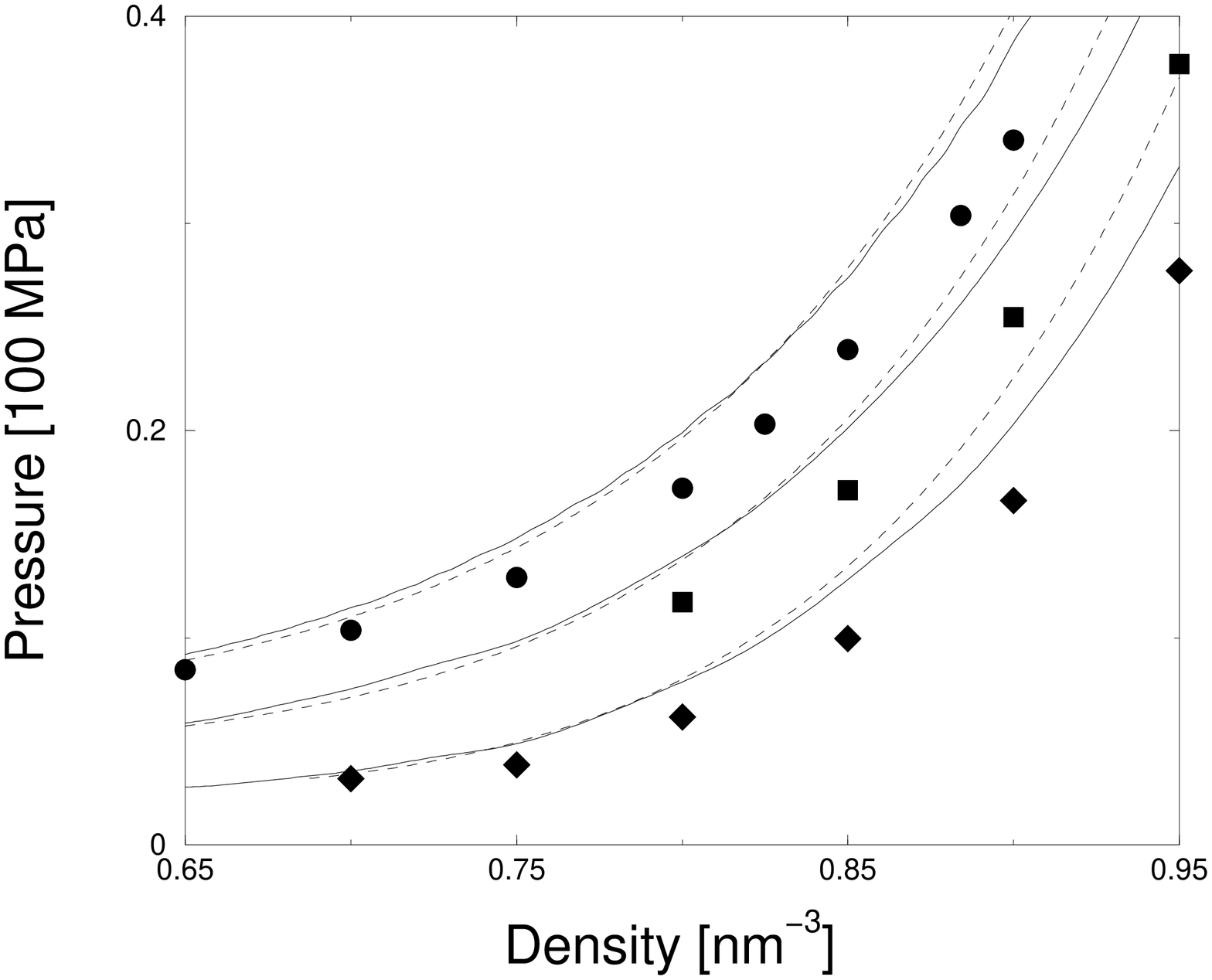}}
\end{center}
\end{minipage}
\begin{minipage}[!t]{\dimen0}
\begin{center}
{\includegraphics[width=6.4cm,angle=-90]{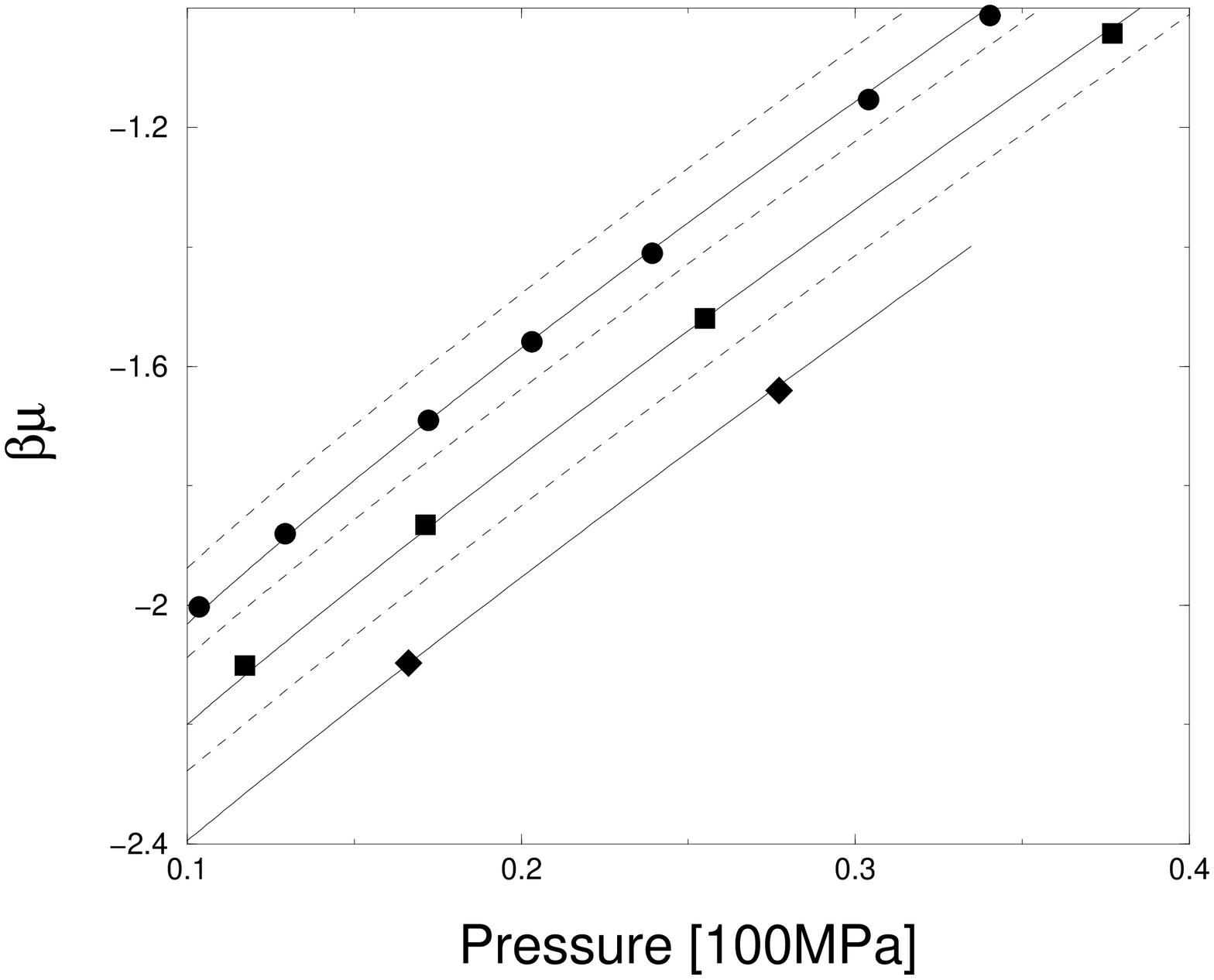}}
{\includegraphics[width=6.5cm,angle=-90]{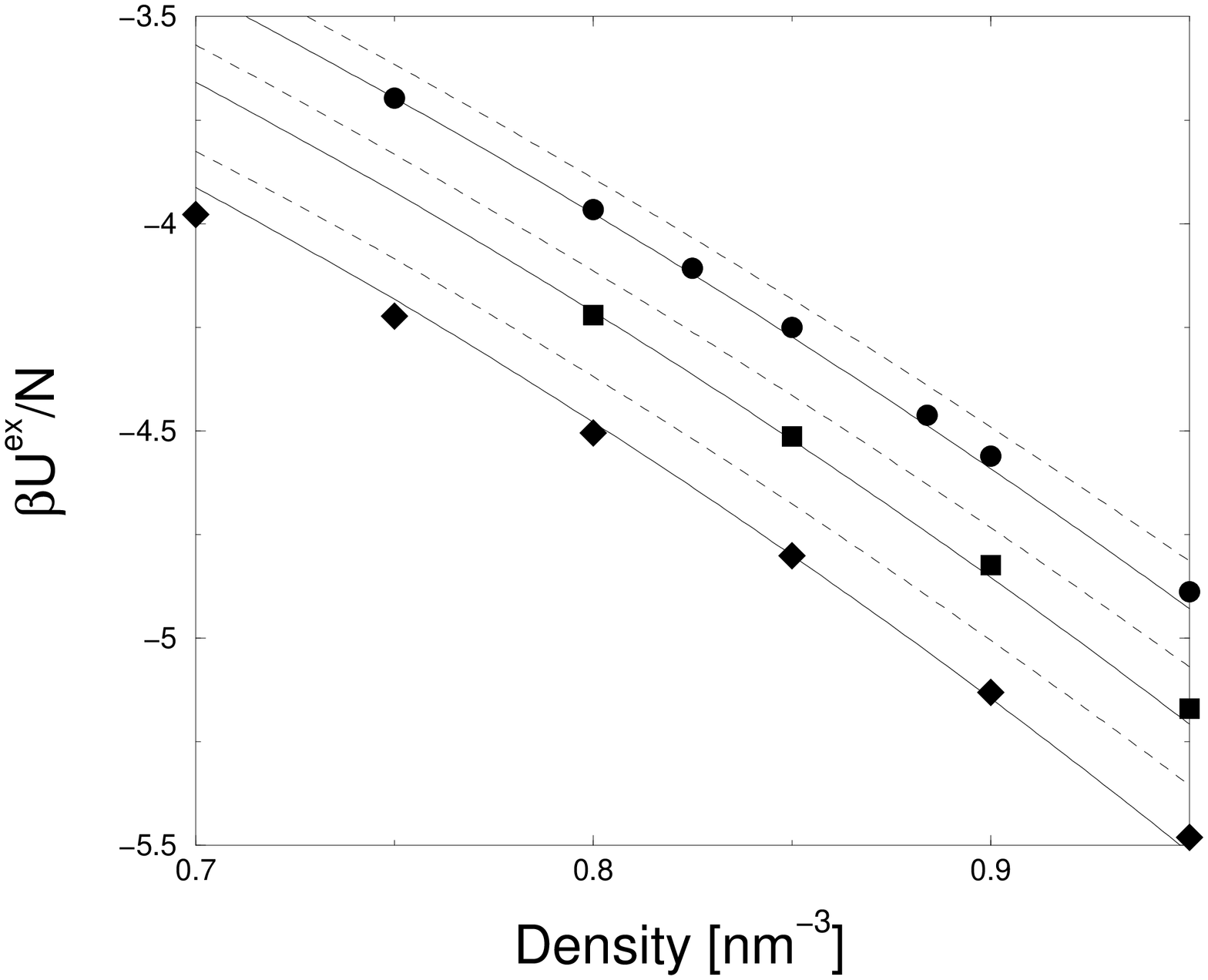}}
\end{center}
\end{minipage}
\caption{MHNC (full lines) and SCOZA (dashed lines) 
thermodynamic properties along three isotherms
in the fluid phase. Symbols:
Monte Carlo results~\protect\cite{noi} at $T=2100$~K (circles),
2000~K (squares) and 1900~K (diamonds). 
All curves are displayed from top to bottom
in order of decreasing temperature.
}\label{fig:freeIE} 
\end{figure*}

\begin{figure*} 
\begin{center} 
\includegraphics[width=9cm,angle=-90]{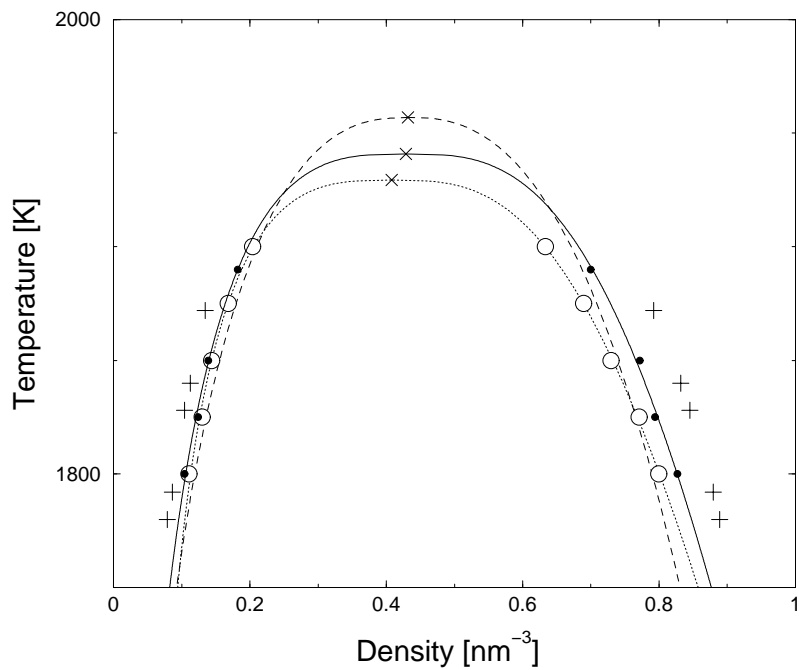} 
\caption{MHNC (dotted line with open circles) and 
SCOZA (dashed line) liquid-vapor
coexistence; the GEMC results 
(full line with dots,~\protect\cite{Fucile}) are also shown.
The crosses are the critical points.
Pluses: MHNC under a local consistency constraint~\protect\cite{CMHNC}.
The calculated coexistence points are shown as
open circles (MHNC) and dots (GEMC) while 
the lines represent corresponding theoretical interpolations.
}\label{fig:binodal} 
\end{center} 
\end{figure*}

\begin{figure*} 
\begin{center} 
\includegraphics[width=9.0cm,angle=-90]{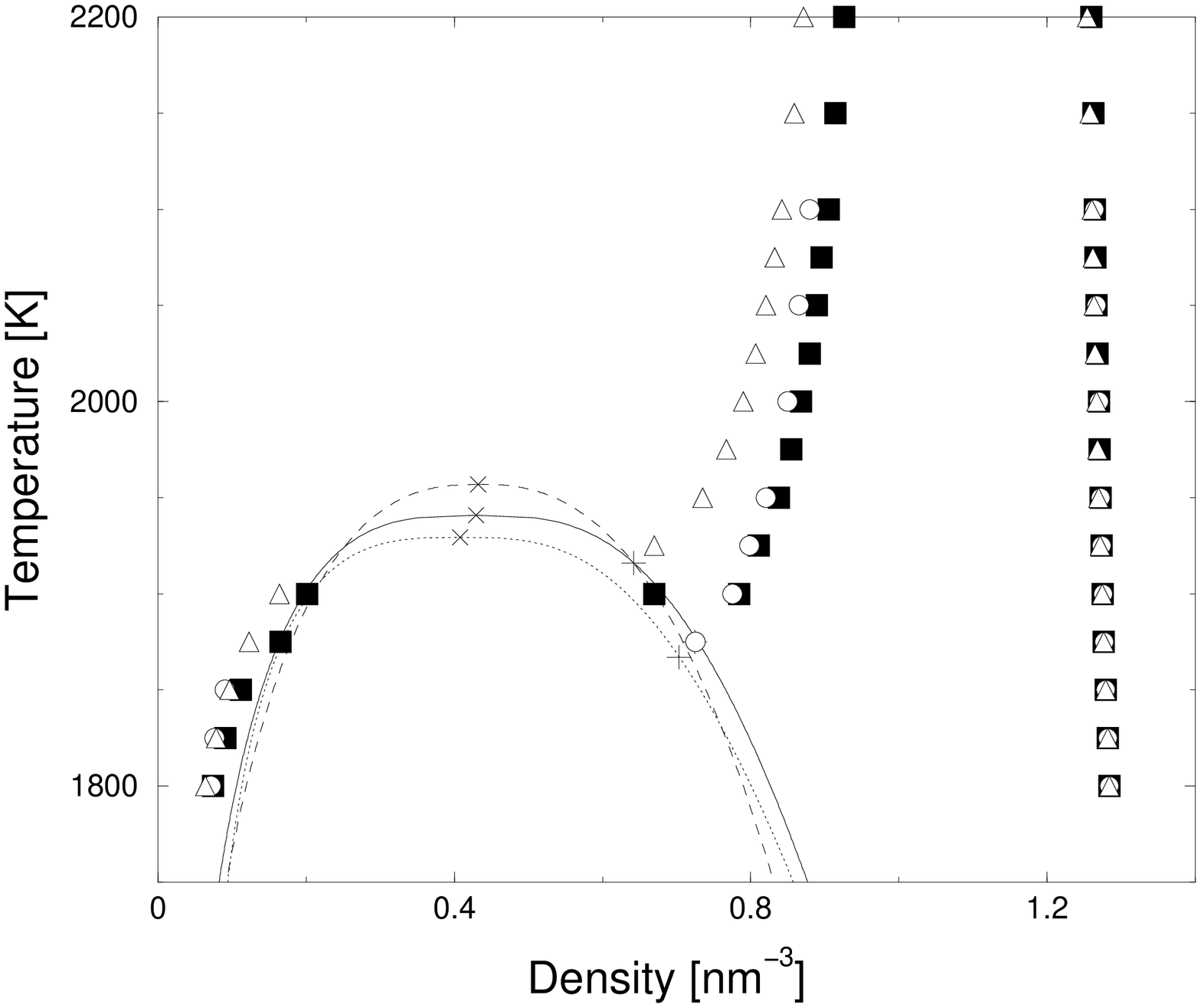}  
\includegraphics[width=9.0cm,angle=-90]{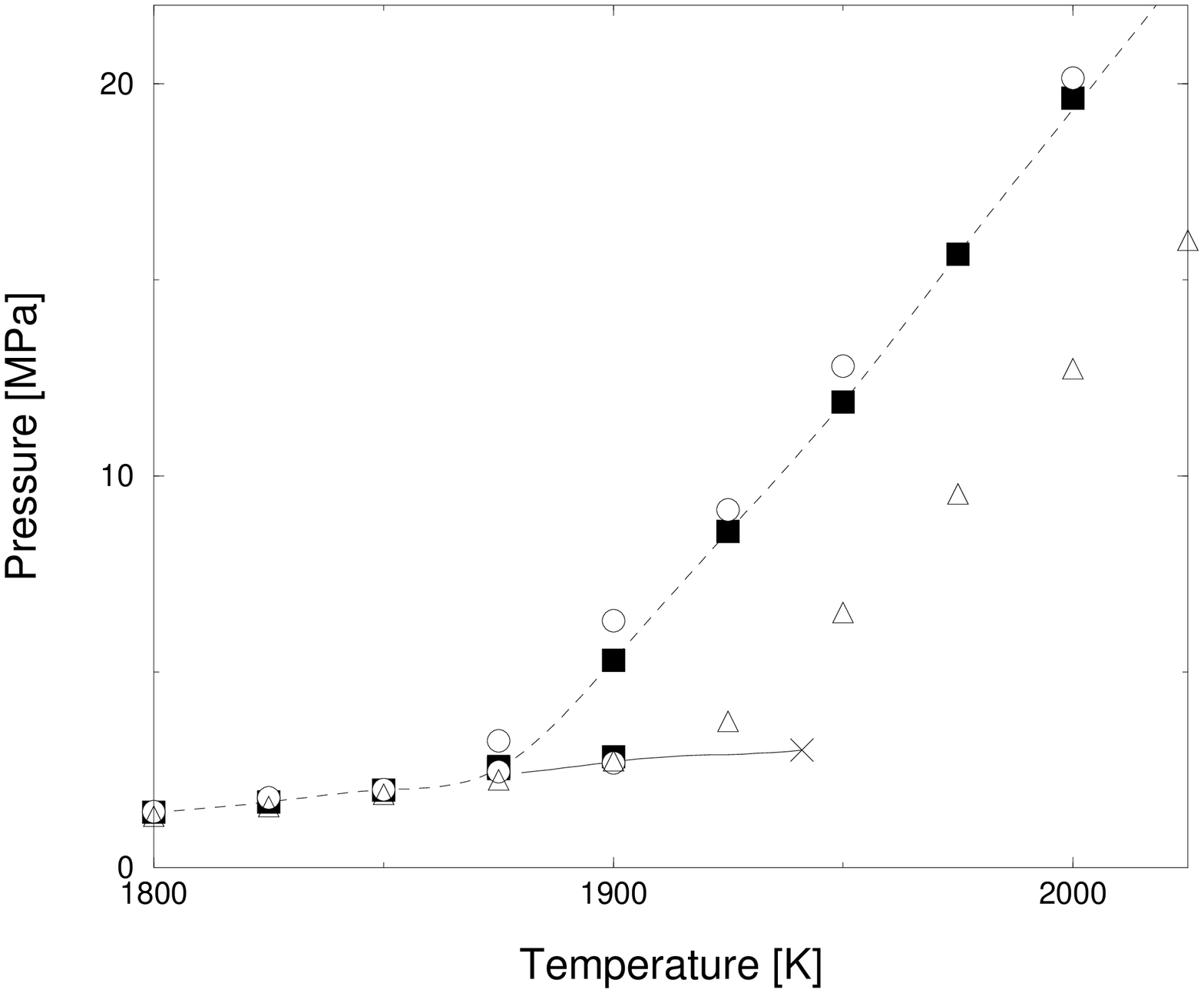} 
\caption{MHNC (circles) and SCOZA (triangles) fluid-solid
coexistence densities (top) and pressures (bottom).
Squares: Monte Carlo results~\protect\cite{noi} 
with the corresponding triple point (plus).
For completeness,
the binodal curves drawn in  Fig.~\ref{fig:binodal} 
are also reported in the top panel.
In the bottom panel the full line is  the GEMC 
liquid-vapor coexistence pressure~\protect\cite{Fucile} with the corresponding
critical point (cross); dashed lines are guides to the eye
for the simulation results.
}\label{fig:phdiaIE} 
\end{center} 
\end{figure*}

\begin{figure*} 
\begin{center} 
\includegraphics[width=9.0cm,angle=-90]{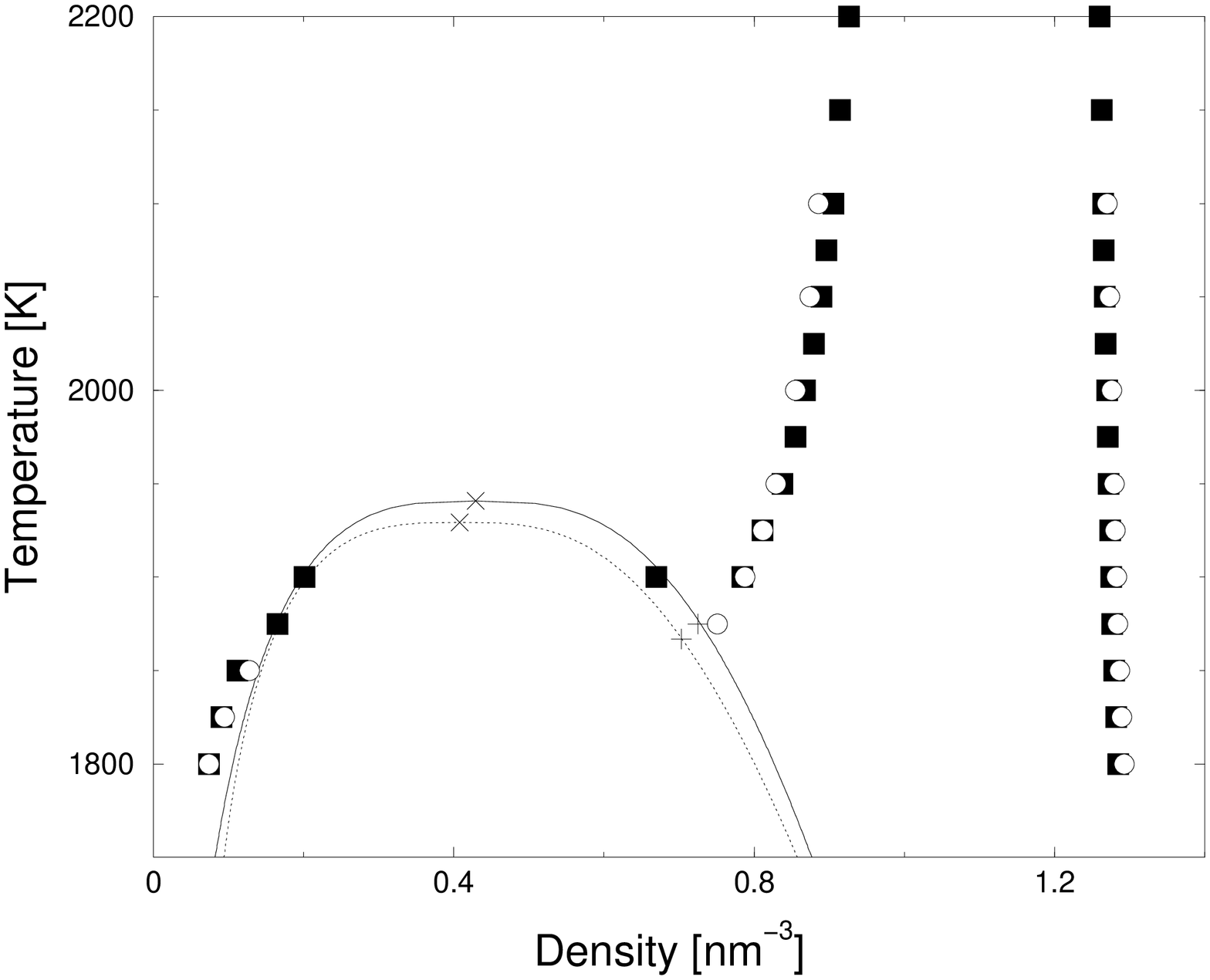}  
\includegraphics[width=9.0cm,angle=-90]{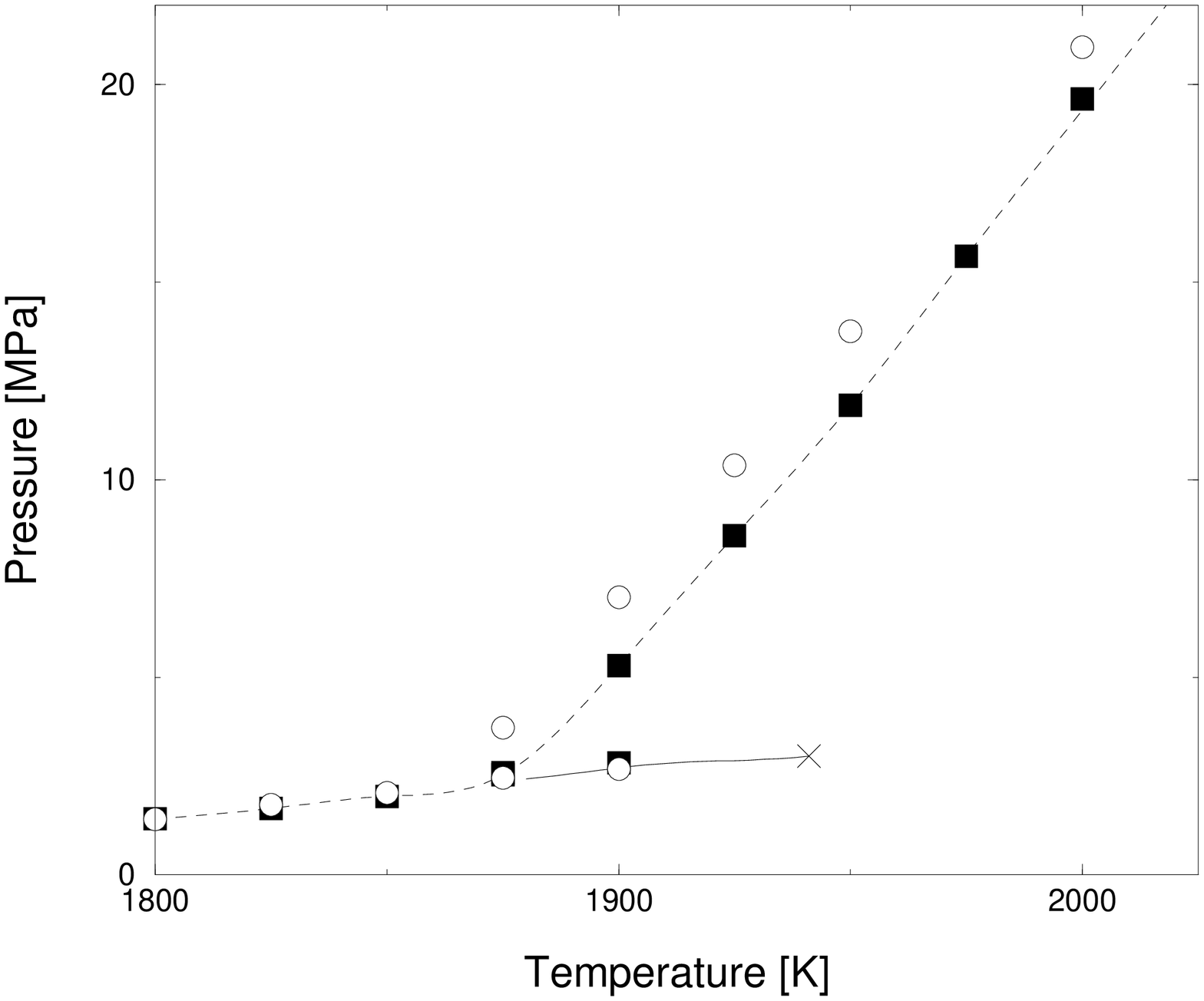} 
\caption{Circles: fully theoretical fluid-solid coexistence 
densities (top) and pressures (bottom) of the \C60 model.
Squares: simulation results~\protect\cite{noi}. 
The MHNC and GEMC binodal curves with corresponding critical points
shown in Fig.~\ref{fig:binodal} are also displayed.
In the bottom panel the full line is  the GEMC 
liquid-vapor coexistence pressure~\protect\cite{Fucile} with the corresponding
critical point (cross); dashed lines are guides to the eye
for the simulation results.
}\label{fig:phdiaTH} 
\end{center} 
\end{figure*}

In the first part of this section 
(Figs.~\ref{fig:freePT}-\ref{fig:phdiaPT})  
we compare the perturbation theory for the solid phase
with the corresponding Monte Carlo results~\cite{noi}. 
Then, the PT free energies are combined  
with the simulation data for the fluid phase in order to
determine the solid-fluid boundaries.
In the second part (Figs.~\ref{fig:freeIE}-\ref{fig:phdiaIE}) we present a 
complementary approach where
the MHNC and SCOZA theories  
are used to predict the liquid-vapor equilibrium, and~---~in combination 
with the simulation inputs for the solid phase~\cite{noi}~---~the full 
phase diagram. 
Hinging on both comparisons, we have calculated 
in the last part (Fig.~\ref{fig:phdiaTH}) 
the phase diagram on the basis of purely
theoretical approaches. 
 
As far as the PT is concerned, 
the expression~(\ref{eq:sigma1}) 
gives a hard-core equivalent diameter
smoothly varying from $\sigbh=0.970$~nm at $T=2200$~K 
to $\sigbh=0.972$~nm at $T=1800$~K. Such a weak variation  
reflects clearly
the stiff, rapidly varying nature of the repulsive 
part of the Girifalco potential. The more refined  
expression~(\ref{eq:sigma}) for the hard-core diameter, $\sigwca$, 
which adds  a further dependence 
on the density, involves just a minor correction 
to $\sigbh$ evaluated through Eq.~(\ref{eq:sigma1}). 
 
We report in  Figs.~\ref{fig:freePT}-\ref{fig:pressPT} 
the PT free energy, chemical potential and   
equation of state along several isotherms in the solid phase,
as obtained  
by using either $\sigwca$ or $\sigbh$
as hard-sphere diameter, together 
with approximations~(\ref{eq:exp}) or~(\ref{eq:exp2}) 
for the free energy. In the figures, the theoretical predictions 
are compared to the corresponding simulation results~\cite{noi}. 
As can be expected,
the use of either $\sigwca$ or $\sigbh$
produces similar results. 
The PT predictions generally agree 
with simulation data all over the solid phase,  
especially in
the chemical potential vs the pressure. 
The equation of state  
compares less favorably to Monte Carlo 
results (see Fig.~\ref{fig:pressPT}),  
although wider discrepancies occur for $\rho \ge 1.27$~nm$^{-3}$,  
i.e. almost outside the coexistence region, since  
the melting line is located  
between $\rho\simeq 1.25$ and $\rho\simeq1.27$~nm$^{-3}$. 
Remarkably, 
the accuracy of the numerical estimates 
in the expression~(\ref{eq:exp})
results from
the balance between 
two almost comparable~---~and relatively large with 
respect to their difference~---~contributions of opposite signs. 
The PT predictions slightly get worse 
if the second order correction to the free energy  
(however small, since it hardly exceeds 2\% of the main contributions) 
is taken into account. 
As observed in the original paper~\cite{BH}, 
this can depend on the semimacroscopic derivation  
of the last term 
in Eq.~(\ref{eq:exp2}), that is more appropriate 
for longer interaction ranges,
in such a way that a reasonably  
large number of particles fits into the region of attractive  
potential. The derivative 
of the pressure involved in Eq.~(\ref{eq:exp2}) can as well represent 
a source of numerical uncertanties.  
 
We may anticipate that the fine reproduction  
of the chemical potential, as obtained through the first-order PT, 
eventually leads to the most accurate predictions 
for the phase behavior of the model. 
For such reason, we concentrate  
on the expansion~(\ref{eq:exp}) and  
show, in Fig.~\ref{fig:phdiaPT}, 
the corresponding phase diagram; 
predictions corresponding both
to $\sigwca$ and to $\sigbh$ are reported.
The coexistence lines are determined
by using the simulation results of Ref.~\cite{noi} for the fluid phase. 
It appears that 
the PT gives accurate predictions 
for the coexisting 
temperatures and densities, with marginal effects 
related to distinct prescriptions for the hard-core diameter. 
By converse,
a systematic overestimate characterizes the coexistence  
pressure above the triple point 
(see bottom panel of Fig.~\ref{fig:phdiaPT}), 
especially if $\sigwca$
is employed. 
We argue in this latter case that  
the definition of the cavity function  
inside the hard core~\cite{Choi},  
directly borrowed from the theory of simple fluids~\cite{HG}, 
is to some extent  unappropriate for the  
\C60 model; moreover, numerical uncertainties 
can occur,  due to the more cumbersome procedure needed to  
evaluate $\sigwca$ through Eq.~(\ref{eq:sigma}), 
which involves in particular the calculation 
of the slope of $\overline g_{\rm hs}(r)$  
at close contact. 
 
In summary,
it turns out
that a PT first-order expansion of the model's free energy around a  
properly defined reference solid gives 
reliable predictions for the coexistence properties, 
and, more generally, for the solid phase behavior of the system at issue. 
This finding is noteworthy in that it implies
that the hard-sphere behavior dominates, to a large extent,
the structure of the solid phase; conversely, as discussed in~\cite{noi},
the properties of the fluid phase
are markedly affected by energetic aspects  
related to the attractive part of the potential. 
Our results further support the use of the PT
to characterize the solid phase
of systems interacting through short-range forces,
already documented in Refs.~\cite{Amokrane,Nava,Foffi}.

As far as the theoretical description of the fluid phase is concerned,  
we report in Fig.~\ref{fig:freeIE}
the SCOZA and MHNC predictions 
for the free energy, the chemical potential, 
the pressure, and the internal energy,
along with
the corresponding Monte Carlo results of Ref.~\cite{noi}. 
It appears that 
the MHNC predicts to a high accuracy 
the chemical potential as a function of the pressure, a quantity 
directly related to the coexistence 
properties of the model. A satisfactory agreement with the  
simulation data also emerges for both the free and  
the internal energies, while 
a marked discrepancy affects the estimate of the 
pressure,  
especially in the high-density regime. 
On the other hand,  as is visible in Fig.~\ref{fig:freeIE},
the thermodynamic properties  of the model
are generally overestimated in the SCOZA framework. 
 
The tendency of the MHNC to
overestimate the pressure under the global consistency procedure,
already observed in Ref.~\cite{Peppe},
appears somehow unexpected, especially considering the overall good 
performances of this theory. 
We can conjecture that it might depend on an overestimate
of the effective hard-sphere diameter $\sighs^*$ entering 
the bridge function as the adjustable 
parameter to enforce the thermodynamic consistency 
in Eq.~(\ref{eq:gc1}) or~(\ref{eq:gc2}). 
Indeed, the value of the virial pressure $P_{\rm v}$
depends on a delicate balance between contributions of opposite signs.
The use 
of the Verlet-Weis bridge functions 
(which fit available simulation data), 
instead of the analytical Percus-Yevick ones, 
might as well improve the predictions for the pressure. 
As for the SCOZA,
we argue that the discrepancies with simulation data
must be related to the approximate treatment of  
the repulsive part of the potential. 
In particular, the substitution of the soft-core interaction  
with a purely hard-sphere potential for $r<r_0$ 
causes an enhanced repulsion (which reflects  
in the overestimate of the free energy and of the pressure), especially 
at high temperature, 
where shorter distances can be sampled by the system. 
We have indeed verified by few simulation runs
along the isotherm $T=2100$~K (not reported here)
that the agreement between SCOZA and Monte Carlo  pressures 
considerably improves, provided that 
both approaches are compared on the same model,
i.e. hard-spheres plus a Girifalco tail.

We report in Fig.~\ref{fig:binodal} the SCOZA and MHNC 
predictions for the liquid-vapor coexistence properties of the model. 
The Gibbs Ensemble Monte Carlo (GEMC) binodal line
obtained in Ref.~\cite{Fucile} with a sample of 1500 particles 
is also shown for comparison.
It comes out that both theories  
reproduce quite faithfully the binodal curve, especially  
as far as the vapor branch is concerned. 
The coexisting liquid densities are slightly underestimated, 
the SCOZA reverting this trend just below the critical point.  
The binodal curve obtained through
the MHNC under the global consistency constraint
definitely improves on
the local one, as displayed in Fig.~\ref{fig:binodal},  
thus confirming
the preliminary evidence reported in Ref.~\cite{Peppe}. 
As for the SCOZA, an internal compensation  
between the overestimate of the chemical potential  
on one hand, and of the pressure on the other hand, 
could be at the origin of the fairly good agreement 
with the GEMC binodal curve.  

\begin{table}[!t]
\begin{center}
\caption{MHNC and SCOZA
critical and triple point densities (in nm$^{-3}$) and
temperatures (in K).
Simulations:
the critical point corresponds to the GEMC estimate
with 1500 particles of Ref.~\protect\cite{Fucile};
the triple point has been calculated in Ref.~\protect\cite{noi}.
}\label{tab:1}
 \begin{tabular}{c|ccc}
 & MHNC  & SCOZA & Simulations\\
 \hline
 $T_{\rm cr}$    &  1929  & 1957  & 1940 \\
 $\rho_{\rm cr}$ &  0.408 & 0.432 & 0.43 \\
 $T_{\rm tr}$    &  1867  & 1916  & 1880   \\
 $\rho_{\rm tr}$ &  0.70 & 0.64   & 0.73
 \end{tabular}
\end{center}
\end{table}

The MHNC solution algorithm converges to a thermodynamic consistent 
solution up to $T=1900$~K (see Fig.~\ref{fig:binodal}).  
In view of early studies 
on HNC-type theories (see e.g.~\cite{report}), the difficulties 
to explore the critical region  
may be intrinsic to such approaches, rather than an artifact of numerical 
procedures. Consequently,
the available MHNC binodal 
points must be fitted to some expected behavior in order to determine
the critical parameters;
as a common practice, we have used 
the scaling law for the coexisting densities to calculate the critical  
temperature, $T_{\rm cr}$:  
\be\label{eq:crit} 
\rho_{\rm liquid} - \rho_{\rm vapor} \propto |T-T_{\rm cr}|^\beta \,, 
\ee 
where the exponent takes on  
the non-classical effective value $\beta=0.32$.  
Once the critical temperature is known,
we have calculated the 
critical density, $\rho_{\rm cr}$, 
by fitting the MHNC binodal points with the law of  
rectilinear diameter:  
$\rho_{\rm liquid} - \rho_{\rm vapor} = \rho_{\rm cr} + A|T-T_{\rm cr}|$, 
with $A$ to be determined from the fit. 
At variance with the MHNC, within the SCOZA
the critical region can be approached in principle 
with 
arbitrary precision  and therefore
neither a fitting procedure, nor a scaling 
behavior hypothesis must be invoked to calculate the 
critical point parameters. 
It emerges from Fig.~\ref{fig:binodal} 
that the theoretical predictions  
closely bracket the GEMC critical point.  
The MHNC, SCOZA, and GEMC critical parameters are  
collected in Table~\ref{tab:1}. 

In Fig.~\ref{fig:phdiaIE} we report 
the fluid-solid phase boundaries, 
as obtained by the combination of the MHNC and SCOZA predictions for the fluid  
phase and the simulation results of Ref.~\cite{noi} for the solid phase;
for a more comprehensive overview,
we also show 
a sketch of the binodal curves just discussed, 
as well as the 
full simulated fluid-solid coexistence points.
As it is apparent,
the overall good quality of the MHNC thermodynamic predictions 
provides a faithful reproduction 
of the phase boundaries of the model. A small overestimate of  
the fluid-solid coexistence 
pressure only occurs above the triple point temperature.  
As for the SCOZA, the solid-vapor equilibrium and the melting line
are satisfactorily
predicted, while the observed shift of the chemical potential 
(see Fig.~\ref{fig:freeIE})
gives rise to a marked 
discrepancy in the coexistence pressure above the triple point; 
on the other hand,
the overestimate of the pressure of the fluid phase  
is reflected in the shift of the theoretical 
freezing line toward lower densities. 
 
We have determined 
the MHNC and SCOZA triple points, also reported in Table~\ref{tab:1}, 
from the intersection between the corresponding binodal 
and freezing lines.  It comes out   
that the MHNC predicts the existence 
of a stable liquid phase 
with  a small shift
of the triple and critical densities to lower values,
with respect to
the simulation results.
The SCOZA also predicts  
a liquid pocket for the model, although  
restricted to a narrower temperature and density range.

It emerges
that the MHNC theory for the fluid phase, and the first-order PT for 
the solid phase represent, among these envisaged here, 
the most accurate theoretical 
tools to enquire into the full phase portrait of the \C60 model. 
The ensuing
phase diagram is reported in Fig.~\ref{fig:phdiaTH}.  
The MHNC binodal line is directly drawn from Fig.~\ref{fig:binodal}. 
As is visible, the agreement with simulation data is semiquantitative; 
the comparison 
of Figs.~\ref{fig:phdiaPT},~\ref{fig:phdiaIE}, and~\ref{fig:phdiaTH}
demonstrates that
the errors in the location of phase boundaries
are essentially brought about by the MHNC, while the use of the PT 
for the solid phase
does not introduce any appreciable discrepancy 
with simulation results. 
For such reason we do not expect a significant
improvement of the overall appearance of the phase diagram
if the SCOZA predictions for the fluid phase substitute
the MHNC ones.
In the bottom panel of Fig.~\ref{fig:phdiaTH},
the overestimate of the fluid-solid coexistence pressure
magnifies slightly the common trend 
already observed for the PT and MHNC approaches separately.
 
\section{Summary and concluding remarks} 
 
We have presented a theoretical investigation of the 
thermodynamic properties and the phase behavior
of the Girifalco \C60 model,
in the framework provided by 
three refined theoretical tools, as
the Modified Hypernetted Chain (MHNC) 
 with a global thermodynamic consistency
constraint and the Self-Consistent Ornstein-Zernike Approximation (SCOZA)
for the fluid phase, and a Perturbation Theory (PT) with various degrees
of refinement for the free
energy of the solid phase.

The theoretical
predictions are assessed against our recent simulation study~\cite{noi}. 
It turns out that the phase portrait of the model
is predicted fairly well, in particular as for the solid-vapor
coexistence and the melting line.
By converse, the inaccuracies in
the thermodynamic predictions sensitively affect
the determination of the fluid-solid coexistence pressure
above the triple point.
The MHNC and SCOZA liquid-vapor binodals are sufficiently
accurate; the critical point estimates 
closely
bracket the Gibbs Ensemble Monte Carlo datum~\cite{Fucile}.
In comparison with the simulation and the MHNC freezing lines
(which turn to be  practically superimposed), 
the SCOZA tends to underestimate
the density of the liquid branch of the fluid-solid coexistence,
with a  corresponding shift of the triple
point to lower densities and higher temperatures.

We have interpreted our results 
in view of the basic assumptions underlying each theory.
It has emerged that the SCOZA
mainly suffers from the substitution of the Girifalco
soft-core repulsion at short distance with a 
purely hard-sphere interaction, which leads to a corresponding overestimate
of the pressure and of the free energy in the fluid phase.
We have also conjectured that the MHNC pressure might improve 
if the Verlet-Weis bridge functions were adopted instead of the
Percus-Yevick ones, as currently adopted in this work. 
As for the PT, it turns out that 
both a second order expansion of the free energy, and
a refined definition
of the equivalent hard-core diameter, 
lead to a slight worsening of the  
theoretical predictions on the whole, which could be related to more
cumbersome numerical procedures involved.

We have eventually succeeded in predicting
to a satisfactory degree of accuracy 
the phase diagram of the Girifalco model, on 
the basis of purely theoretical approaches.
We have selected for this purpose 
a sophisticated MHNC treatment of the fluid phase,
and the first-order PT for the solid phase.
We expect that our conclusions 
about the most suitable theoretical schemes 
to investigate 
the phase diagram of the Girifalco model,
can be reasonably extended to other systems characterized
by similar interparticle interaction laws.
In particular, we plan  to analyze
the phase diagram
of other higher-order C$_{n>60}$ fullerenes,
combining
both simulations
and theoretical schemes.
We have already produced
several Monte Carlo results for the
Girifalco C$_{84}$ fullerene, to be used
for such purpose. The related calculations 
are currently in progress.

\acknowledgments

ESP and GK acknowledge the financial support by the \"Osterreichische
Forschungsfonds (FWF) under Project Nos. P14371-TPH and P15758-TPH.

\newpage

\end{document}